\documentstyle[12pt,a4,graphicx]{article}

\newcommand{\be}{\begin{equation}}
\newcommand{\ee}{\end{equation}}
\newcommand{\ba}{\begin{eqnarray}}
\newcommand{\ea}{\end{eqnarray}}
\newcommand{\dis}{\displaystyle}

\begin{document}
\begin{titlepage}
\begin{flushright}
CAFPE-105/08\\
UG-FT-235/08
\end{flushright}
\vspace{2cm}
\begin{center}

{\large\bf Obtaining $\sigma \to \gamma \gamma$ Width
from Nucleon Polarizabilities
\footnote{
Work supported in part 
by MICINN, Spain and FEDER, European Commission (EC)
Grant Nos. FPA2005-01678 (J.B.), FPA2006-05294 (J.P.), 
Consolider-Ingenio 2010 Grant No. CSD2007-00042 -- CPAN,
by Junta de Andaluc\'{\i}a Grants No. P05-FQM 101 (J.P.),
P05-FQM 467 (J.P.) and P07-FQM 03048  (J.P.) and
 by the EC RTN network FLAVIAnet Contract 
No.  MRTN-CT-2006-035482 (J.P.).
Invited talk given by J.P.
at ``14$^{\rm th}$ International Conference on Quantum Chromodynamics
 (QCD '08)'', 7-12 July 2008, Montpellier, France.}}\\

\vfill
{\bf Joaquim Prades$^{a)}$ and Jos\'e Bernab\'eu$^{a)}$}\\[0.5cm]

$^{a)}$ Centro Andaluz de F\'{\i}sica de las Part\'{\i}culas
Elementales (CAFPE) and Departamento de
 F\'{\i}sica Te\'orica y del Cosmos, Universidad de Granada \\
Campus de Fuente Nueva, E-18002 Granada, Spain.\\[0.5cm]

$^{b)}$  Departament de F\'{\i}sica
Te\`orica, IFIC, Universitat de Val\`encia-CSIC,\\
Apt. de Correus 22085, E-46071 Val\`encia, Spain.\\[0.5cm]

\end{center}
\vfill
\begin{abstract}
 We propose a new method that fixes the coupling to two photons
of the recently found lightest QCD resonance, the $\sigma$.
This coupling provides crucial information for discriminating
the yet unknown nature of this special state. Our method uses 
available data on the nucleon polarizabilities together
with analyticity and unitarity. Taking into account all the
uncertainties, our result is $\Gamma_{\rm pole}= 1.2 \pm 0.4 $ keV. 
\end{abstract}
\vfill
September 2008
\end{titlepage}

\section{Introduction}

The  lowest isospin $I=0$ and angular
momentum $J=0$ QCD resonance  is usually
called the $\sigma$  and  plays a special role in the QCD dynamics
and in the QCD non-perturbative vacuum structure. Recently,
it has been shown that the $\sigma$ is also the lowest QCD resonance by
fixing the mass and width of this state with a precision
of just tens of MeV in Ref. \cite{CGL06}. 
Making an analytic continuation of the $I=0$ and 
$J=0$ partial wave S-matrix in the region
of validity of  Roy equations, these authors  find a zero
at $E=[(441^{+16}_{-8} - i \, (272^{+9}_{-12})]$ 
MeV on the first Riemann sheet which reflects the 
$\sigma$ pole on the second Riemann sheet.
Also on the first Riemann sheet, the inverse
of the partial wave $\pi\pi$ S-matrix
$S=1\,+\, 2 i \, \beta(t) \, T(t)$ has a zero at $E^*$.
Here, 
\be
\label{T}
T(t)=\frac{1}{\beta(t) \cot[\delta(t)]+ \beta(t)}\, ,  
\ee
$\delta(t)$ is the scalar-isoscalar $\pi\pi$ phase shift, 
$\beta(t)=\sqrt{1- 4 m_\pi^2 / t}$  and $t=E^2$. The position
of the $\sigma$ resonance pole has been confirmed
in \cite{GPY07} at $E=[(484\pm 17 - i \, (255\pm 10)]$
MeV.

  The nature of the $\sigma$ remains one of the most
intriguing and difficult issues in particle physics.
There are have been many proposals about its substructure:
 $\overline{q} - q$ state,  $\pi - \pi$ molecule, 
$(\overline{qq}) - (qq)$ tetraquark,  glueball, and of
course, several admixtures of these substructures. 
The size of $\sigma \to \gamma \gamma$ width 
can shed  light on this question. 

 $\gamma \gamma \to (\pi\pi)_{I=0,2}$
amplitudes have been calculated using twice-subtracted 
dispersion relations in \cite{PEN06} and \cite{ORS08}
including the recent data on $\pi\pi$ final state interactions
 which contain
the $\sigma$ pole in the scalar-isoscalar contribution.
They get $4.09\pm 0.29$ keV \cite{PEN06}  and $1.68\pm 0.15$ keV 
\cite{ORS08}  for the $\sigma$ into two photons width.
The origin of this discrepancy is discussed in \cite{ORS08}.
More recently, the authors of \cite{Belle08} made
 an amplitude analysis of the world  published data
on $\gamma \gamma \to \pi^+ \pi^-$ and find  two regions of solutions. 
The width $\sigma \to \gamma \gamma$
in these regions are predicted in Ref. \cite{Belle08} 
to be $3.1\pm0.5$ and $2.4\pm0.4$ keV, respectively.

\vspace*{-0.1cm}
\section{Method}

In Ref. \cite{BP08}, we proposed a new method that fixes
the coupling  to two photons  $g_{\sigma\gamma\gamma}$ 
of the $\sigma$ meson found in the
$\pi\pi$ scattering amplitude \cite{CGL06,GPY07} 
using only available precise experimental
data on the nucleon electromagnetic polarizabilities
together with analyticity and unitarity.
This differs from the analysis of \cite{SCH07}, where
the properties of the $\sigma$ meson of a Nambu--Jona-Lasinio
model were used. 
Nucleon electric $\alpha$ and magnetic $\beta$ polarizabilities
are well measured using low energy
Compton scattering on protons and neutrons with  $\alpha+\beta$ 
constrained  by a  forward dispersion relation \cite{DG70}.
 The results are \cite{PDG08}:
$\alpha^{\rm expt}=12.0\ 0.6$ and $\beta^{\rm expt}=1.9\mp0.5$
for protons and $\alpha^{\rm expt}=11.6\ 1.5$ and 
$\beta^{\rm expt}=3.7\mp2.0$ for neutrons. 
Here and in the rest of the paper, polarizabilities are given in 
$10^{-4}$ fm$^3$ units.

The authors of Ref. \cite{BEF74,BT77} wrote a sum rule for
$\alpha-\beta$ using a backward dispersion relation for the physical 
spin-averaged amplitude. 
The $s$-channel part of this sum rule 
is related to the multipole content of the
total photo-absorption cross-section.
While the $t$-channel part is related through a dispersion relation 
to the imaginary part of the
amplitude  which using unitarity
is given by the processes $\gamma \gamma \rightarrow
\pi \pi$  and $\pi \pi \rightarrow \overline N N$ 
 \cite{BT77}. The result  is the  BEFT sum rule, 
\ba
(\alpha-\beta) &=&  \frac{\dis 1}{\dis 2 \pi^2} \, 
 \int^\infty_{\nu_{\rm th}}
\, \frac{{\rm d} \nu}{\nu^2} \, \sqrt{1+2 \frac{\nu}{M_p}}
\left[ \sigma(\Delta \pi = {\rm yes}) - \sigma(\Delta \pi = {\rm no}) 
\right] \nonumber \\ &+& \int^\infty_{4 m_\pi^2}
\, \frac{{\rm d} t}{4 M_p^2 -t} \, \frac{\beta(t)}{t^2} \, 
\Big\{ |f_+^0(t)||F_0^0(t)| \nonumber \\
&-&   \frac{(4 M_p^2-t)(t- 4 m_\pi^2)}{16} \, 
|f_+^2(t)||F_0^2(t)| \Big\} , 
\label{BEFT}
\ea
where $M_p$ is the proton mass, the  partial wave helicity
amplitudes $f_+^0(t)$ and $f_+^2(t)$ for 
$\overline N N \rightarrow \pi \pi$ are  Frazer and Fulco's \cite{FF60},
and the partial wave helicity amplitudes $F_0^0(t)$ and $F_0^2(t)$
for $\gamma \gamma \to \pi \pi$ were defined in \cite{BAB76}.
The $s$-channel part of the integrand is obtained  from that of
the forward physical amplitude by changing the sign of the non-parity
flip multipoles $(\Delta \pi= {\rm no}$) and yields
$(\alpha -\beta)^s = -(5.0\pm 1.0)$ \cite{SCH05}. 
The ``experimental'' $(\alpha-\beta)^t$ is therefore
$15.1\pm 1.3$ for protons and $12.9\pm2.7$ for neutrons.
Products of helicity amplitudes in  Eq.
(\ref{BEFT}) appear only as moduli products, which can be 
negative  if the phases differ from the $\pi\pi$
phase shift in an odd number of $\pi$'s.
 The d-wave contribution to $(\alpha-\beta)^t$ is much smaller
than the s-wave one; hence, it is a good approximation
to take just the Born  result 
$(\alpha-\beta)^t_2 \simeq -1.7$ \cite{HN94}. Finally,
the ``experimental'' value for $(\alpha-\beta)^t_0$
is $16.8\pm 1.3$ for protons and $14.6\pm 2.7$ for neutrons.
The input $|F_0^0(t)|$ for $(\alpha-\beta)^t_0$  in Eq. (\ref{BEFT}) 
is what we want to fix using this  ``experimental'' value. 
The other input for this quantity, 
the Frazer-Fulco's $|f_+^0(t)|$ amplitude, is 
 known with enough accuracy for our purposes from the old determination
in \cite{BOH76}, 
though  it could be improved using recent data on $\pi \pi$
phase shift. We have assigned a 20 \% to the theoretical determination 
of $(\alpha-\beta)^t_0$ from this source. Notice that the  $1/t^2$ factor
in the integrand of $(\alpha-\beta)^t_0$ makes the well known
low-energy, and to a lesser extent, intermediate-energy contributions
to be the dominant ones. 

On the physical sheet, we can write the twice-subtracted dispersion 
relation  \cite{DISP},
\be
F_0^0(t) \, =  L(t)   - \, \Omega(t)  \left[ c \, t
+ \, \frac{t^2}{\pi}
\int_{4 m_\pi^2}^{\infty} \, \frac{{\rm d} t'}{t^{'2}}
\frac{L(t') {\rm Im} \, \Omega^{-1}(t')}{t'-t-i \varepsilon} \right]
\,   
\label{DEFA}                
\ee
where $c$ is a subtraction constant fixed by
chiral perturbation theory (CHPT) \cite{DISP,GL85}, 
$c= \alpha /48 \pi f_\pi^2$ with 
$\alpha \simeq 1/137$ the fine-structure constant,
$f_\pi=92.4$ MeV the pion decay constant,
 \be
\Omega(t) = {\rm exp} \left[ \frac{t}{\pi} \int_{4 m_\pi^2}^\infty
\, \frac{{\rm d} t'}{t'} \, \frac{\delta(t')}{t'-t-i \varepsilon} \right]
\ee
is the scalar-isoscalar $\pi\pi$ Omn\`es function \cite{OMN58}
which gives the correct right-hand cut contribution 
and $L(t)$ is the left-hand cut contribution.
In this way we ensure unitarity, the correct analytic structure
of $F_0^0(t)$ and that the $\sigma$ pole properties 
enter through  the scalar-isoscalar 
phase-shift $\delta(t)$ from $T(t)$ in (\ref{T}). 
Here, we shall use a simple analytic
expression for $T(t)$, compatible with Roy's equations, which takes a
three parameter fit from \cite{GPY07} 
including both low energy kaon data and 
high energy data. This fit is valid
up to values of $t$ of the order of 1 GeV$^2$, which is enough for the
$(\alpha-\beta)^t_0$ integrand in Eq. (\ref{BEFT}).
  
At the $\sigma$ pole  position
on the first Riemann sheet \cite{PEN06,ORS08}
\be
\label{defggamma}
F_0^0(t_\sigma) \, =  \,  e^2 \, \sqrt{6} \, 
 \frac{g_{\sigma\gamma\gamma}}{g_{\sigma\pi\pi}}\, 
 \frac{1}{2 i \beta(t_\sigma)},
\ee 
where $e$ is the electron charge, 
$g_{\sigma\pi\pi}^2$ 
is the residue of the $\pi\pi$ scattering amplitude at the $\sigma$ pole
on the second Riemann sheet and $g_{\sigma\gamma\gamma} \, 
g_{\sigma \pi\pi}$ is proportional to the residue of the $\gamma \gamma
\to \pi\pi$ scalar-isoscalar scattering amplitude on the second
Riemann sheet. The proportionality factors are such that
$g_{\sigma\pi\pi}$  
and $g_{\sigma\gamma\gamma}$ agree with those  used  in 
\cite{PEN06}. The  pole width is given by 
\be
\label{width}
\Gamma_{\rm pole}(\sigma \to  \gamma \gamma) = 
\frac{\alpha^2 | \beta(t_\sigma) \, 
g_{\sigma\gamma\gamma}^2|}{4 M_\sigma} \, 
\ee
that agrees, modulo normalizations,  with that of Ref. \cite{ORS08}.

Low's theorem fixes the amplitude $F_0^0(t)$ 
to be  the Born term at very low-energy \cite{LOW}.
As   first approximation, we therefore consider  the left-hand
cut contribution $L(t)$ in (\ref{DEFA}) to be the Born 
contribution $L_B(t)$, 
\be
L_B(t)= e^2 \frac{1-\beta(t)^2}{\beta(t)} \log \left( 
\frac{1+\beta(t)}{1-\beta(t)} \right) \, .
\ee

\begin{figure}[Ht]
\begin{center}
\includegraphics[width=3in,height=2in]{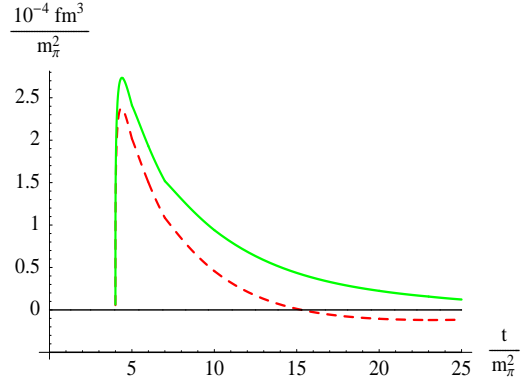}
\caption[pilf]{\protect \small The integrand of $(\alpha-\beta)^t_0$
in (\ref{BEFT}). The dashed line is when using $L(t)=L_B(t)$  in 
(\ref{DEFA}) and 
the continuous line is  when using $L(t)=L_B(t)+L_A(t)+L_V(t)$ 
in (\ref{DEFA}) as explained in the text.
\label{figure}}
\end{center}
\end{figure}

This leads to $F_0^0(t)|_B$ which when input 
 in the sum rule (\ref{BEFT}) gives
$(\alpha-\beta)^t_0|_B=6.7 \pm 1.2$, where the uncertainty comes
mainly from the input data on $|f_+^0(t)|$. This result
is 5$\sigma$ away the ``experimental'' values quoted above.
When  analytically continued to complex $t$, the amplitude $F_0^0(t)|_B$  
has a pole  at $t_\sigma = \left\{ \left[(474 \pm 6) -i (254\pm 4)\right] 
{\rm MeV} \right\}^2$
with $g_{\sigma \pi \pi}=\left[ (452\pm4) + i (224\pm2) \right]$ MeV
and  using (\ref{defggamma}) we get 
$g_{\sigma\gamma\gamma}/g_{\sigma\pi\pi}|_B=
(0.49^{+0.03}_{-0.01}) -i \, (0.37\pm 0.03)$ which leads to
$\Gamma_{\rm pole}(\sigma \to \gamma \gamma)|_B= 2.5 \pm 0.2$ keV.
But this $F_0^0(t)|_B$ does not reproduce the ``experimental''
$(\alpha-\beta)^t_0$ and hence
we need to go beyond the Born approximation $L_B(t)$ for the left-hand cut.

 The first corrections to $L_B(t)$  originate
in the resonance exchange $\gamma \pi \to  R \to \gamma \pi$, 
with  $R=a_1, \rho$ and  $\omega$ \cite{ORS08,DISP}.
In the  narrow width approximation, the $a_1$ exchange 
contribution to $L(t)$ is
\be
\label{LA}
L_A(t) =   
e^2 \frac{C}{32 \pi f_\pi^2} \left[ 
t + \frac{M_{a_1}^2}{\beta(t)} \log \left( \frac{1+\beta(t)+t_A/t}
{1-\beta(t)+t_A/t} \right) \right] \,  
\ee
while the $\rho$ and $\omega$ resonances exchange contribution to $L(t)$  
in nonet symmetry  ($M_\rho=M_\omega=M_V \simeq$ 782 MeV) is
\be
L_V(t)=\nonumber \\ 
e^2  \frac{4}{3} R_V^2   \left[ 
t - \frac{M_{V}^2}{\beta(t)} \log \left( \frac{1+\beta(t)+t_V/t}
{1-\beta(t)+t_V/t} \right) \right] \, 
\ee
 with $t_R = 2(M_R^2-m_\pi^2)$.
 The low energy limit of $L_V(t)$  goes as $t^2$  and 
we fix $R_V^2=$ 1.49 GeV$^{-2}$ from  the well known
$\omega \to \pi  \gamma$ decay. Though the  low energy  limit of $L_A(t)$  
goes as $t$  and  corresponds to the charged 
pion electromagnetic polarizability 
 $(\overline \alpha-\overline \beta)_{\pi^\pm}$
or equivalently to $L_9 + L_{10} = (1.4 \pm 0.3) \cdot 10^{-3}$
in CHPT \cite{BC88}, we 
 consider $L_A(t)$ as an effective contribution  for moderate higher values
of $t$ with $C$ a real constant 
to be determined phenomenologically and not connected to the pion
polarizability. This is supported by the fact that  the
 $a_1  \pi \gamma$  interaction is not so well known at 
intermediate energies. We fix the effective $C$  by requiring that 
the ``experimental'' value  of $(\alpha - \beta)^t_0$ is reproduced
within 1.5 standard deviations  of the total uncertainty
when $L(t)$ in (\ref{DEFA}) is given by $L(t)=L_B(t)+L_A(t)+L_V(t)$. 
This procedure leads to $C = 0.59\pm 0.20$  
and the integrand of the sum rule is 
given in Fig. \ref{figure} as a continuous line. 
Notice that the zero at $t_0$  in the integrand  of $(\alpha-\beta)^t_0$
in (\ref{BEFT})
when using $F_0^0(t)|_B$ has clearly 
disappeared now. 

The low-energy $\gamma \gamma \to \pi^0 \pi^0$ cross-sections
obtained when the left-hand cut is either $L_B(t)$ or
the full $L(t)$ case studied  above are  similar \cite{ORS08}.
The central values are compatible 
with data for values of $t$ below 
$(450 \, \, {\rm MeV})^2$  and are above data 
but compatible   within two standard deviations 
for larger values of $t$   up to 
$(600 \, \, {\rm MeV})^2$  and  within one  standard deviation for $t$ 
between  $(600 \, \, {\rm MeV})^2$ and $(800 \, \, {\rm MeV})^2$. 

\section{Results and Conclusions}

  The scalar-isoscalar amplitude 
$F_0^0(t)$, obtained using $L(t)=L_B(t)+L_A(t)+L_V(t)$
as explained above,  is analytically continued to the complex plane,
  and at $t_\sigma$ on the first Riemann sheet one gets 
$g_{\sigma\gamma\gamma}/g_{\sigma\pi\pi}=
(0.23^{+0.05}_{-0.09}) \, - \, i \, (0.30 \pm 0.03)$
which has  a smaller absolute value when compared
with  ${g_{\sigma\gamma\gamma}/g_{\sigma\pi\pi}}{|}_B$  
 and leads to $\Gamma_{\rm pole }(\sigma \to \gamma \gamma)= 
(1.0\pm 0.3 )$ keV.  
The error quoted here is from the uncertainties in  
the ``experimental'' 
value of  $(\alpha-\beta)^t_0$ and the inputs of the 
sum rule (\ref{BEFT}) only.
To obtain the rest of the uncertainty, 
we vary the $\sigma$ properties in the $\pi\pi$ scattering
as follows. We still use the three parameter fit formula including 
low energy kaon data  and high energy  data for $\cot(\delta(t))$ 
in  \cite{GPY07}  as input in the amplitude $T(t)$
but with parameter values slightly
modified to reproduce the $\sigma$ pole position 
$t_\sigma= ([(441\pm 6)  - i \, (272\pm 4)]\,  {\rm MeV})^2$  
found in  \cite{CGL06}. In this case, 
we get $g_{\sigma\pi\pi}= [(480\pm 7) \, + \, i\, 
(191\pm 3)]$ MeV. 
 With that $T(t)$ input in the dressed Born amplitude in Eq. (\ref{DEFA}),
one gets $(\alpha-\beta)^t_0{|}_B=6.1 \pm 1.1$,
${g_{\sigma\gamma\gamma}/g_{\sigma\pi\pi}}{|}_B=
(0.57 \pm 0.02) - i (0.41 \pm 0.03)$  and    
$\Gamma_{\rm pole}(\sigma \to \gamma \gamma){|_B}=(3.8\pm0.4)$ keV.
The integrand of $(\alpha-\beta)^t_0$ in
(\ref{BEFT}) for this case is very similar to the dashed
line of Fig. \ref{figure}. The effective value of $C$ in 
(\ref{LA}) is $C=0.62\pm 0.20$
when fixed to reproduce the ``experimental'' value of 
$(\alpha-\beta)^t_0$ within 1.5 standard deviations of the total
uncertainty.
With this new $C$, the analytic continuation to  $t_\sigma$
gives  $g_{\sigma\gamma\gamma}/g_{\sigma\pi\pi}=
(0.31^{+0.05}_{-0.07})\,  - \, i \, (0.32\pm0.03)$ and 
$\Gamma_{\rm pole}(\sigma \to \gamma \gamma)=(1.5\pm0.4)$ keV. 
Again, the integrand of $(\alpha-\beta)^t_0$ in
(\ref{BEFT}) for this case is very similar to the
 continuous curve of Fig. \ref{figure}. 

Making a weighted average of
the two cases  discussed above,  we get
\be
\Gamma_{\rm pole} (\sigma \to \gamma \gamma) = (1.2\pm0.4)
\, \, {\rm keV} \, .
\label{final}
\ee

\section*{Acknowledgments}
It is a pleasure to thank Stephan Narison for the invitation
to this very enjoyable conference. 
We also thank Heiri Leutwyler, Jos\'e A. Oller,
Jos\'e Ram\'on Pel\'aez and Mike Pennington for useful
discussions and sharing unpublished results.

\end{document}